\documentclass[12pt]{article}
\usepackage{epsfig}

\newcommand{\bra}[1]{\langle  {#1}  \vert }
\newcommand{\ket}[1]{\vert {#1} \rangle }

\begin {document}
\begin{flushright}
{\small
SLAC--PUB--9047\\
November 2001\\}
\end{flushright}

\vfill

\begin{center}
{{\bf\LARGE Moving Mirrors, Black Holes, Hawking}
\vskip 20pt
{\bf\LARGE Radiation and
All That $\ldots$}\footnote{Work supported by Department
of Energy contract DE--AC03--76SF00515.}}

\bigskip
{\it Marvin Weinstein \\
Stanford Linear Accelerator Center \\
Stanford University, Stanford, California 94309 \\
E-mail:  niv@slac.stanford.edu}
\medskip
\end{center}

\vfill

\begin{center}
{\bf\large
Abstract }
\end{center}

In this talk I show how to
canonically quantize a massless scalar field in the 
background of a Schwarzschild black hole in Lema\^itre coordinates and
then present a simplified derivation of Hawking radiation based upon
this procedure.
The key result of quantization procedure is that 
the Hamiltonian of the system is explicitly time dependent and so 
problem is intrinsically non-static.
From this it follows that, although a unitary time-development operator exists,
it is not useful to talk about vacuum states; rather, one should
focus attention on steady state phenomena such as the
Hawking radiation.  In order to clarify the approximations used
to study this problem I begin by discussing the related
problem of the massless scalar field theory calculated in the
presence of a moving mirror.

\vfill

\begin{center}
{\it Presented at the Workshop on \\
Light-cone physics: particles and
strings, TRENTO 2001\\
September 3 -- September 11, 2001 }\\
\end{center}

\vfill \newpage

\section{Introduction}

The phenomenon of Hawking radiation has fascinated physicists
ever since Hawking's 1974 paper\cite{hawking} showed that, when
quantum fields are taken into account,
a black hole of mass $M$ appears to emit (nearly) thermal radiation with a temperature 
$T_{\rm H} = 1/(8\pi G M)$.  This result, extensively studied in the literature
(see for example \cite{unruh,jacobson,parikh}), is clearly
robust since all approaches lead to the same conclusion.
So far as I know, however, no derivation of Hawking
radiation discusses the problem within the framework of canonical quantization.
This is one reason the question of whether the time evolution of
the system is unitary became a topic of debate.

In this talk I present work\cite{bhlett} done in collaboration with Kirill Melnikov,
showing that a simple canonical quantization procedure exists and
leads to the usual Hawking results within the framework of a unitary
quantum theory.  This theory has some surprising
features but doesn't seem lead to paradoxes which invalidate the approach.
One unique advantage of our calculation is that we can compute the energy-momentum
tensor at all positions and times.  This allows us to explicitly
exhibit transient behavior and retardation effects, and --
as a matter of principle -- give a self-consistent treatment 
of back reaction.

Before diving into details I will spend a few moments reminding you why
carrying out a Hamiltonian formulation of the problem
of a massless scalar field in the presence of a black-hole background
seems so difficult.  Consider the usual Schwarzschild metric for a
black hole of mass $M$:
\begin{equation}
	ds^2 = -(1-{2M\over r})\,dt^2 + {1\over (1- {2M\over r})}\,dr^2 + r^2 d\Omega^2
\label{schwarz}
\end{equation}
and a massless scalar field with Lagrange density
\begin{equation}
	{\cal L} = \sqrt{-g} \left[ g^{\mu \nu} \,\partial_\mu \phi(x)\,
	 \partial_\nu \phi(x) \right] .
\end{equation}
(Note that in what follows I will set $2M = 1$ to simplify the equations.)
While one usually focuses on the singularity of the metric
at $r=2M$, this coordinate singularity is not a problem.  The real 
issue is that in order to canonically quantize this theory we need a family of
spacelike slices which foliate the spacetime.  Given this, we can then define 
the field and its conjugate momentum on one of these slices.
Inspection of the metric, Eq.\ref{schwarz}, shows that surfaces of constant
Schwarzschild time change from spacelike to timelike at the horizon ($r=2M$)
and so, they do not fulfill our requirements.  

\section{Summary of Results}

In order to define a set of spacelike surfaces which
foliate the Schwarzschild spacetime we begin by introducing Painlev\'e coordinatess
because surfaces of constant Painlev\'e time extend from $r=0$ to $r=\infty$ and are
everywhere spacelike.  Since, however, the
Painlev\'e timelines are not everywhere timelike we will have to make one more transformation,
to Lema\^itre coordinates, in order to canonically quantize the theory.
In these coordinates we see that the metric and therefore the Hamiltonian
is explicitly time dependent, which is the reason why Hawking radiation exists.
Of course, despite the time dependence of the Hamiltonian, the theory possesses a
unitary time development operator, which is all we need to discuss the relevant
physical issues.  Note, because the Hamiltonian is time dependent, it is never
useful to talk about eigenstates.

The strategy of the computations I will describe goes as follows: first, we canonically
quantize the theory; second, we devise a scheme (the Geometric Optics
Approximation) for approximately solving the Heisenberg equations of motion
(this is easier than calculating the unitary time development operator); third, we
compute the temperature seen by a thermometer located at fixed $r$ which is adiabatically
switched on and off; last, we compute the flux passing through a sphere of fixed
radius and show that in the long time limit (after transients have died out)
it has the Hawking value.

\section{Review of Coordinate Systems}

While we are ultimately interested in the surfaces defined by fixing Painlev\'e time,
it is convenient to first introduce Kruskal coordinates and plot the various surfaces
of interest in these coordinates.  The reason for this is that these coordinates
make it particularly easy to draw null-geodesics (they are simply lines parallel to
either the $X$ or $Y$ axes shown in Fig.1) and they allow us to easily compare
surfaces of fixed Schwarzschild time to surfaces of fixed Painlev\'e time.
Kruskal coordinates are defined by the equations:
\begin{equation}
 x y = (r-1)\,e^{r} \qquad , \qquad
		{x \over \vert y \vert} = e^{t_S} .
\label{kruskaleq}
\end{equation}
In these coordinates the Schwarzschild metric becomes:
\begin{equation}
 ds^2 = {32\,  e^{\vert -r \vert} dx\,dy \over r} + r^2 d\Omega^2
\end{equation}
Note that Eq.\ref{kruskaleq} immediately shows us that fixed Schwarzschild $r$ is a
hyperbola in the $x,y$-plane, as shown in Fig.1, and that surfaces of fixed
Schwarzschild time corresponds to straight lines $x=\vert y \vert\,e^{t_S}$ (which are not
shown in Fig.1).

Now, Painlev\'e coordinates are arrived at by making an $r$-dependent shift in
Schwarzschild time; i.e.,
\begin{equation}
 t = \lambda - 2\sqrt{r} - \ln\left(\left\vert{\sqrt{r} - 1 \over \sqrt{r} + 1}\right\vert\right)
\end{equation}
These are the almost horizontal curves shown in Fig.1 which clearly foliate
the spacetime. Note, that while Schwarzschild $t$ and Painlev\'e $\lambda$
differ by functions of $r$, if two events having the same $r$ then the differences in
Schwarzschild and Painlev\'e times are equal.
In Painlev\'e coordinates the Schwarzschild metric takes the form
\begin{equation}
ds^2 = -(1 - {1 \over r}) \,d\lambda^2 + {2\,d\lambda\,dr \over \sqrt{r}} + dr^2 +  r^2 d\Omega^2
\end{equation}
\epsfverbosetrue
\begin{figure}
\begin{center}
\leavevmode
\vskip 1in
\epsfig{file=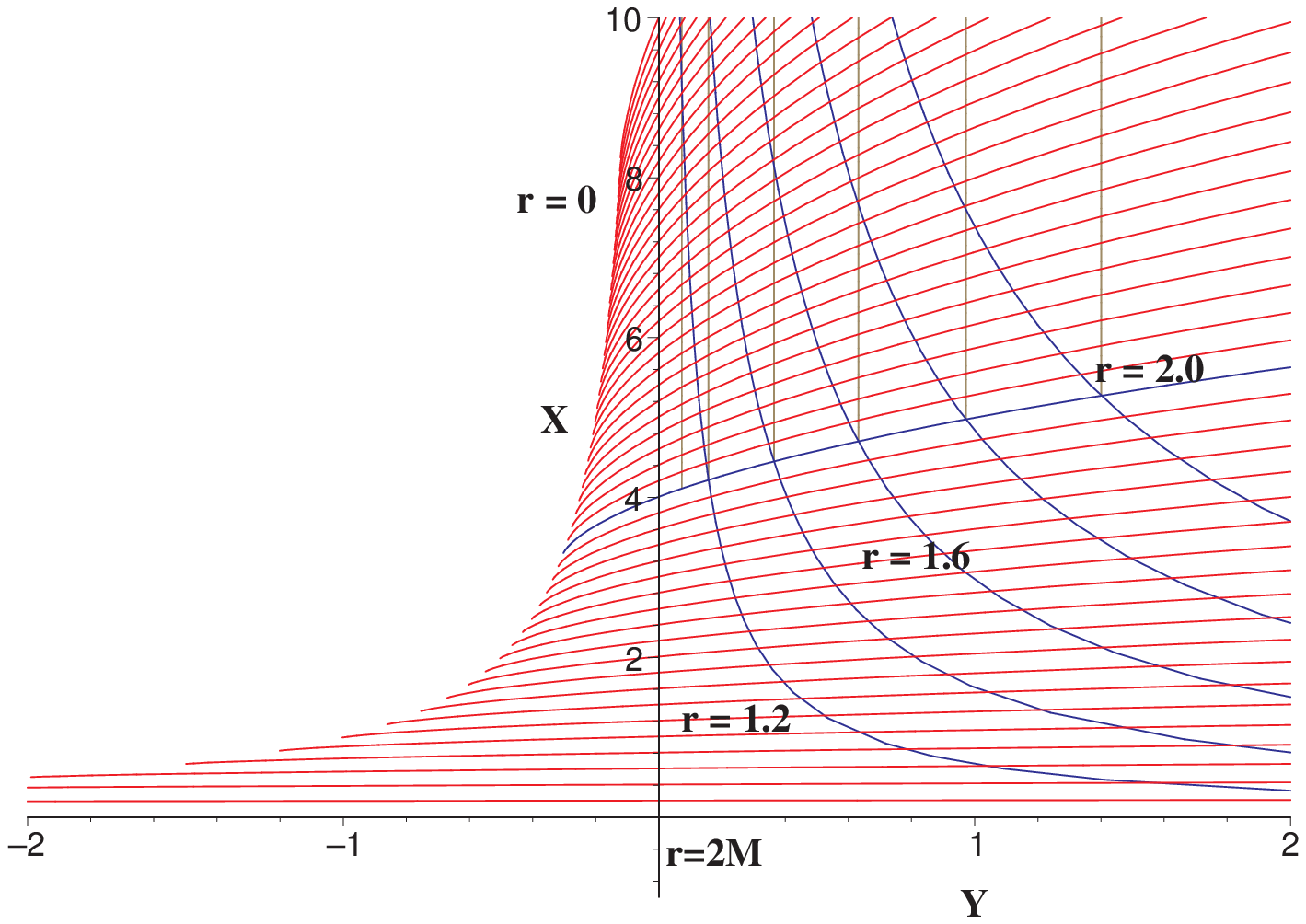,width=4.5in}
\end{center}
\vskip -1.75in
\caption[kruskaletc]{}
\label{kruskal}
\end{figure}
Although Painlev\'e coordinates are useful for defining our set of spacelike surfaces
they don't work well for canonical quantization because of the cross term
involving $d\lambda\,dr$ which appears in the metric.  A better coordinate system
is provided by Lema\^itre coordinates, which are related to Painlev\'e $\lambda$ and $r$ by
\begin{equation}
	r(\lambda,r_{sch}) = (r_{sch}^{3/2} - {3\over 2}\lambda)^{2/3}
	= ({3\over 2} (\eta - \lambda))^{2/3}
\end{equation}
In Lema\^itre coordinates the metric takes the form
\begin{equation}
ds^2 = - d\lambda^2 + {1\over r(\lambda,\eta)} d\eta^2 + r(\lambda,\eta)^2 d\Omega^2
\end{equation}
which allows a straightforward treatment of canonical quantization.

\section{Canonical Quantization}

We choose the surface $\lambda = 0$ as the initial surface on which to canonically
quantize the free massless scalar field theory.
Since the metric in all coordinate systems is rotationally invariant
we can always solve the field equations for each angular momentum mode
separately.  For this reason we can imagine expanding the field in
spherical harmonics in $\theta$ and $\phi$ and then restricting attention
just to the $L=0$ mode, since this contains most of the interesting physics.
If we do this then, in Lema\^itre coordinates, the $L=0$ scalar field Lagrangian
reduces to
\begin{equation}
{\cal L} = \sqrt{-g} \,{1\over 2} \left[ (\partial_\lambda \phi\,(\lambda,\eta))^2
- r\,(\partial_\eta\phi\,(\lambda,\eta))^2 \right]
\end{equation}
where the determinant $\sqrt{-g}$ is
\begin{equation}
\sqrt{-g} = r^{3/2} = {3\over 2} (\eta-\lambda).
\end{equation}
Following the usual rules the momentum conjugate to the field is
\begin{equation}
\pi(\lambda,\eta) =  (\eta - \lambda)\,\partial_{\lambda}\phi(\lambda,\eta ) ,
\end{equation}
and the canonical Hamiltonian is
\begin{equation}
H(\lambda) = {1\over 2} \int_{\lambda}^\infty d\eta\,
\left( {\pi^2 \over \eta - \lambda} + r (\eta-\lambda) (\partial_\eta\phi)^2
\right) .
\end{equation}
The commutation relations for $\phi$ and $\pi$ are
\begin{equation}
\left[ \pi(\lambda,\eta), \phi(\lambda,\eta') \right] = -i\,\delta(\eta - \eta')
\end{equation}
As advertised, this Hamiltonian is explicitly time dependent and
in a sense this finishes our job, since we see that we are looking for
steady state and not static behavior.  
Another feature of this Hamiltonian is that setting $\lambda = 0$ we
see that by a simple rescaling of $\phi$ and $\pi$, in order to absorb the factor of
$1/\eta$ in the $\pi^2$ term, we can
convert it to the usual Hamiltonian of the $L=0$ mode of a free massless field
in flat space and therefore solve it exactly.

It is important to note, as in the usual interaction representation,
the fact that we have a time
dependent Hamiltonian doesn't mean that we don't have a unitary time-development
operator.  There is a one parameter family $U(\lambda)$ which satisfies the equation
\begin{equation}
     {d \over d\lambda} U(\lambda) = H(\lambda)\,U(\lambda)
\end{equation}
whose solution is the path ordered exponential of the integral of $H(\lambda)$.
Fields at later $\lambda$ are defined by
\begin{eqnarray}
\phi(\lambda,\eta) &=& U(\lambda)\, \phi(\eta) \,U^{\dagger}(\lambda) \\
\pi(\lambda,\eta)  &=& U(\lambda) \, \pi(\eta) \, U^{\dagger}(\lambda) 
\end{eqnarray}
It follows from the canonical commutation relations that
these operators satisfy Heisenberg equations of motion of the form
\begin{equation}
\partial_{\lambda} \left[ (\eta - \lambda) \partial_{\lambda}\phi \right]
-\partial_{\eta}\left[ (\eta - \lambda)\,r\,\partial_{\eta}\phi \right] = 0
\end{equation}

\section{Solving The Heisenberg Equations: An Aside}

To explain the Geometric Optics Approximation
to these Heisenberg equations it is convenient
to first discuss these equations in flat space.
I begin with the case where there are no boundary conditions and then
consider the problem of a moving mirror, whose physics is much
closer to that of the black hole.  By a moving mirror, I mean
a free field theory in flat space together with the boundary 
condition that the field vanishes on and to the left of a curve $x(t)$.

To solve the free field Euler-Lagrange equation when there are no boundary conditions
we rewrite the equations as
\begin{eqnarray}
(\partial_t^2 - \partial_x^2 )\,\phi(t,x) &=& 0 \nonumber \\
(\partial_t - \partial_x)\,(\partial_t + \partial_x)\,\phi(t,x) &=& 0
\end{eqnarray}
The general solution to this equation is 
\begin{equation}
\phi(t,x) = f(x-t) + g(x+t)
\end{equation}
where the functions $f$ and $g$ are determined by the values of $\phi(t,x)$ and
its time derivative at $t=0$; i.e.,
\begin{eqnarray}
\partial_t \phi(t=0,x_0) &=& - \partial_x f(x_0) + \partial_x g(x_0) \nonumber \\
\partial_x \phi(t=0,x_0) &=& \partial_x f(x_0) + \partial_x g(x_0)
\end{eqnarray} 
which says that 
\begin{eqnarray}
\partial_x f(x_0) &=& {1\over 2} (\partial_x\phi(x_0) - \pi(x_0)) \nonumber \\
\partial_x g(x_0) &=& {1\over 2} (\partial_x\phi(x_0) + \pi(x_0)) 
\end{eqnarray}

Now consider, as shown in Fig.2, the case of a field theory with moving
boundary $x(t) = -t + A\,(1-e^{-2\,\kappa t})$, where I have chosen to plot
the curve for $A=1$ and $\kappa=1/2$.  In this case the Euler-Lagrange equations
remain unchanged, however the solution needs to be modified to maintain the
boundary condition which says that $\phi(t,x(t)) = 0$.
This is easily done by the trick of adding a reflected
wave, $g_0(x-t)$, so that the general solution has the form
\begin{eqnarray}
\phi(t,x) &=& \theta(x-t)\,f(x-t) + g(t+x) \nonumber \\
& &+ \theta(t-x)\,g_0(x-t)
\end{eqnarray}
We will see that the crucial feature of this solution is that, if one sits at a fixed
point $x$, the reflected rays contributing to $\phi(t,x)$ for $t >> x$
all come from very near the point $x = A$.

\epsfverbosetrue
\begin{figure}
\begin{center}
\leavevmode
\vskip 1in
\epsfig{file=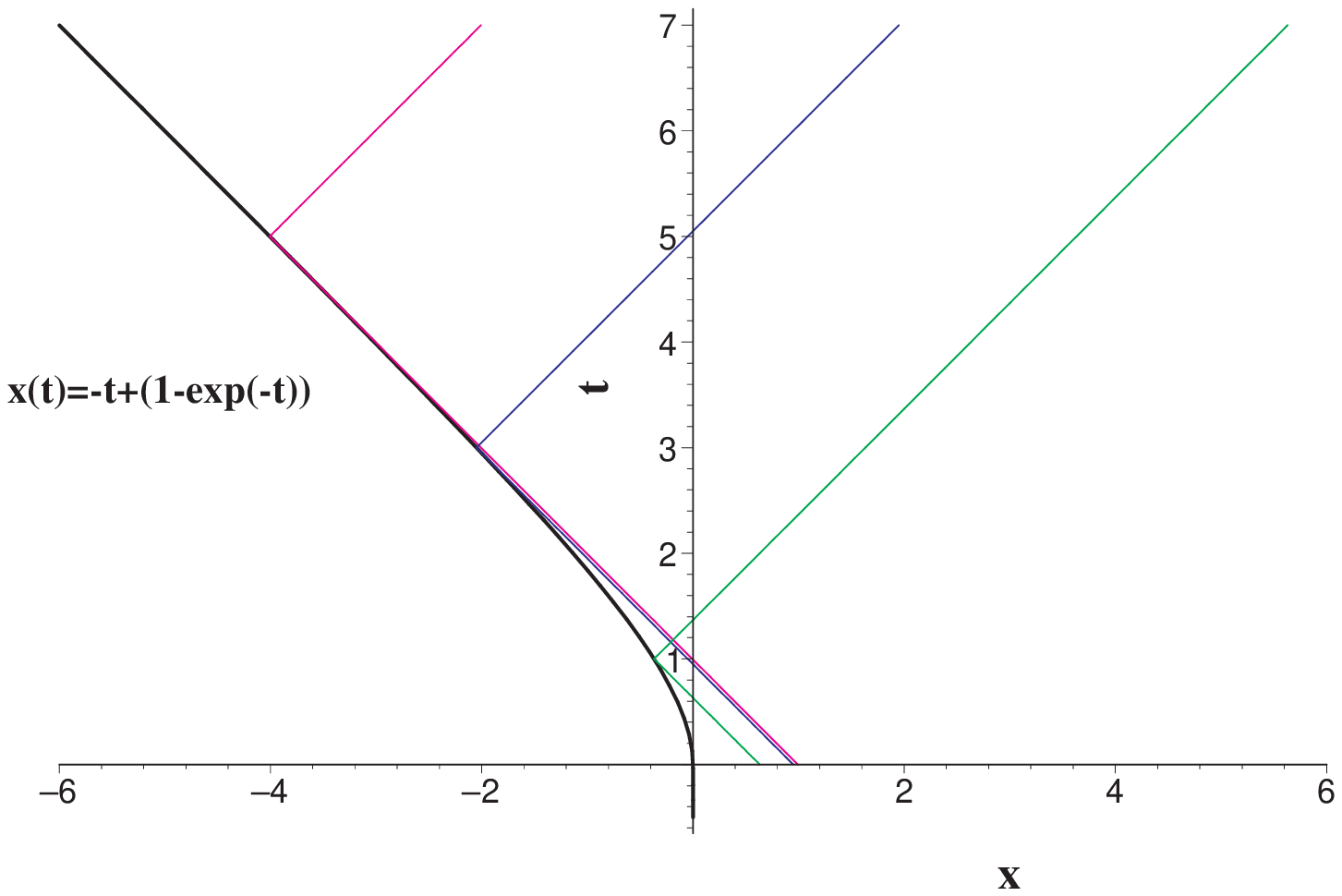,width=4.5in}
\end{center}
\vskip -1.5in
\caption[movingmirror]{}
\label{movingmirror}
\end{figure}

\section{The Story of Temperature}

In what follows I adopt Unruh's definition of a thermometer; i.e., a
quantum system with multiple energy levels interacting with
the field $\phi(t,x)$.  In other words,
I add an interaction of the form 
\begin{equation}
V_{\rm int}(t) = \epsilon\,e^{-(t-t_0)^2/2\sigma} Q\,\phi(x,t)    
\end{equation}
to the Lagrangian,
where the parameters $t_0$ and $\sigma$ define the range in $t$ for which
the interaction is turned on and $x$ specifies the spatial location of the thermometer.
Furthermore, the operator $Q$ is some operator causing transitions among the
energy eigenstates of the thermometer.

Of course, assumptions have to be made in order to get reasonable results.
First, in order for the thermometer to know the mirror is moving, it is
necessary that $t_0 >> x$.  Second, we impose an adiabatic condition,
$\sqrt{\sigma} >> 1/E$, where $E$
is the typical excitation energy of the thermometer, so that we do not excite
the termometer just by turning it on or off.  Finally, we impose the condition
$ E \sim \kappa $, so that the
acceleration of the mirror is capable of exciting the higher states of the thermometer.

With these assumptions second order perturbation theory in $\epsilon$ tells us that
the probability of the thermometer being excited to a state with energy $E$ is
\begin{eqnarray}
{\cal P}(E,E_0) &=& \epsilon^2 \vert \bra{E} M(0) \ket{E_0} \vert^2 
\int dt\,dt' \nonumber\\
&&e^{-i(E-E_0)(t-t')}  e^{-(t-t_0)^2/2\sigma} \nonumber\\
&&e^{-(t'-t_0)^2/2\sigma} \langle \phi(t,x) \, \phi(t',x)\rangle 
\end{eqnarray}
What we now do is compute $\langle \phi(t,x) \, \phi(t',x)\rangle$
by plugging in the formula giving $\phi(t,x)$ in terms of $\phi(x)$ and $\pi(x)$
on the original surface, expand these $t=0$ operators in
terms of annihilation and creation operators
and evaluate the resulting expression.  The key feature of this
problem is that the points on the original surface corresponding to
$(t,x)$ and $(t',x)$ for $t,t' >> x$ are exponentially close to the point
$x=A$.  A straightforward computation gives the result
\begin{eqnarray}
{\cal P}(E,E_0) &=&  {\epsilon^2 \sqrt{\pi} \over 2}\,
{\vert \bra{E} M(0) \ket{E_0} \vert^2 \over E-E_0} \nonumber\\
\times &&\left[ {1 \over e^{2\pi\,(E-E_0)/\kappa} -1} \,\right]
\end{eqnarray}

\section{Computing The Energy Flux}

To compute the flux of energy through a plane at position $x$ we need to compute $T_{tx}$,
the energy momentum tensor for the massless field, which is given by
\begin{equation}
T_{\mu \nu} = {1 \over 2} \left\{ \partial_\mu \phi(t,x), \partial_\nu \phi(t,x) \right\} ,
\end{equation}
Generally, the expectation value of components of $T_{\mu \nu}$ will be divergent
since the field derivatives are being evaluated at the same spacetime point.
To deal with this we adopt a point splitting procedure; i.e., we define
\begin{equation}
T_{tx} = {1 \over 2} \left\{ \pi(t+\delta,x), \partial_x \phi(t-\delta,x) \right\} ,
\end{equation}
evaluate the expression and then take the limit $\delta \rightarrow 0$.
The result is that the energy density $T_{tt}$ diverges
as $1/\delta^2$ but the flux, $T_{tx}$, is finite and unique.
The result of  the computation is
\begin{equation}
T_{tx} = {\kappa^2 \over 48 \pi}
\end{equation}
Although the finiteness of the flux may seem surprising, it is
in fact a consequence of a general theorem.

\section{The Black Hole}

Now let us turn to a discussion of the case of a Schwarzschild
black hole.  Another way of describing the previous example is to say that we
solve the Heisenberg equations by tracing back the two null-rays leaving the point
$(t,x)$ to find the two points at which they intersect the $t=0$ surface and 
write the field in terms of the $\phi$ and $\pi$ at those two points.
There is a simple generalization of this approach to the equations in curved space.

Straightforward analysis shows that what is usually
referred to as a WKB analysis of the wave equation in curved space amounts
to, what we will call, the Geometric Optics Approximation.  The generalization is
given by the following prescription. First, starting from the point 
$(\lambda,r)$, find the two null-geodesics
which meet at this point and trace them back to the surface $\lambda= 0 $
(since this is a geometrical statement it can be solved in any
coordinate system).  In Painlev\'e coordinates the equations for these
two geodesics are
\begin{eqnarray}
	S_1(x_1) &=& \lambda + S_1(r)  \nonumber\\
	S_2(x_2) &=&  \lambda + S_2(r) \nonumber\\
	S_1(r) &=& r - 2\sqrt{r} + \ln((\sqrt{r}+1)^2) \nonumber\\
	S_2(r) &=& -r -2\sqrt{r} - \ln((\sqrt{r}+1)^2) 
\end{eqnarray}
Next, having found the points $x_1$ and $x_2$, write the field at $(\lambda,r)$ as
\begin{equation}
	\phi(\lambda,r) = {1 \over r} \left( \phi_1 (\lambda + S_1(r))
	+ \phi_2 ( \lambda+S_2(r) ) \right)
\end{equation}
where the field on the initial surface is given in terms of two functions
\begin{eqnarray}
\phi(0,r) &=& {1 \over r} \left( \phi_1 ( S_1(r)) + \phi_2(S_2(r)) \right)\nonumber\\
&=& {1 \over r} \left( f_1(r) + f_2(r) \right)
\end{eqnarray}
Finally, in analogy with the flat space case, write $f_1$ and $f_2$ in terms
of $\phi(r)$ and $\pi(r)$ on the surface of quantization ($\lambda=0$).

\section{Black Hole: Thermometer Redux}

Let us now consider adiabatically switching on a thermometer,
kept at fixed Schwarzschild $r$, and then switching it off.  As in the case of the
moving mirror, second order perturbation says
\begin{eqnarray}
{\cal P}(E,E_0) &=& \epsilon^2 \vert \bra{E} M(0) \ket{E_0} \vert^2 \int d\lambda\,d\lambda'\nonumber\\
&&e^{-i(E-E_0)(\lambda-\lambda')} e^{-(\lambda-\lambda_0)^2/2\sigma} \nonumber\\
&&e^{-(\lambda'-\lambda_0)^2/2\sigma} \langle  \phi(\lambda,r) \, \phi(\lambda',r) \rangle
\end{eqnarray}
Of course, now the points $(\lambda,r)$ and $(\lambda',r)$ in the correlation function
$\langle \phi(\lambda,r) \phi(\lambda',r) \rangle$ are traced back, using null-geodesics, to
the initial surface $\lambda=0$ and the fields are appropriately re-expressed
in terms of $\phi$ and $\pi$ on that surface.  The only subtlety in this calculation
is that for arbitrary $r$ the interaction term gets an extra correction for
the time dilation at point $r$, since the energy levels of the thermometer are defined
in its rest frame.  The result of this calculation is that the thermometer reads
a temperature
\begin{equation}
k_B\, T = {1 \over 8 \pi M \sqrt{1 - 2M/r}}
\end{equation}  
which agrees with Hawking's result.

\section{Black Hole: Energy Flux}

The energy flux for the Schwarzschild black hole is calculated in the same way
as for the moving mirror.  Again we have to point-split
the fields appearing in the energy momentum tensor and then take the limit
of zero splitting. As before we find that the flux $T_{\lambda,r}$ is finite
and the total flux through a sphere at large $r$ takes the expected limiting value  
\begin{equation}
  {\rm Flux} = {\pi \over 12}\ T^2  = {\pi\over 12}\ {1 \over (8 \pi M)^2}
\end{equation}
plus, of course, transients which vanish for large $\lambda$ and terms
which die faster than $1/r^2$.

\section{Back Reaction}

Our approach is unique in that we start 
at a finite time and calculate everything as a function
for all values of $\lambda$ and $r$.  Thus, in principle, one can discuss the problem
of back reaction after the Hawking radiation has set in, but before
any appreciable amount of the black hole's mass has been radiated.
The reason there is a back reaction problem is because we calculated the
energy-momentrum tensor for a Schwarzschild
background and found a non-vanishing $T_{\lambda r}$ and so we are in the
situation that 
\begin{eqnarray}
G_{\mu \nu} &=& R_{\mu \nu} - g_{\mu \nu}\,R = 0  \nonumber\\
		T_{\mu \nu} &\neq& 0 .
\end{eqnarray}
This, of course, is not consistent with the Einstein equations and so we do not have
a self-consistent semi-classical problem.

Given our computational procedure, however, for any point $r$, $T_{\lambda r}$ is zero
until radiation from the horizon has a chance to reach that point, at which time an observer
begins to see the Hawking radiation.  In principle we could feed our expression
for $T_{\mu \nu}$ back into the Einstein equations and attempt
to find a self-consistent metric for which the computation of $T_{\mu \nu}$
wouldn't change very much.
Clearly this is difficult to do in general but we can ask what things
look like inside of a sphere of radius $r$ after the Hawking radiation has set in.
If, in this region we adopt a metric of Schwarzschild form but
change $M$ to $M(t)=M_0-F t$ then the metric becomes
\begin{eqnarray}
	ds^2 &=& -(1-{2M(t)\over r})\,dt^2 + {1\over (1- {2M(t)\over r})}\,dr^2 \nonumber\\
	&& + r^2 d\Omega^2  
\end{eqnarray}
which leads to an Einstein tensor of the form 
\begin{equation}
\pmatrix{
0 & {2\,F \over r^2 X(t)} &0 & 0 \cr
{2\,F \over r^2\,X(t)} & 0 & 0 & 0 \cr
0 & 0 & {4\,F^2 \over X(t)^3} & 0 \cr
0 & 0 & 0 & {4\,F^2\sin(\theta)^2 \over X(t)^3} \cr
}
\end{equation}
where
\begin{equation}
X(t) = 1-{ 2\,(M_0 - F\,t) \over r}
\end{equation}
which matches the outgoing flux computed for the static background.
Since our computation
of the flux at a point $(t,r)$ only involves the computation of geodesics
leaving the initial surface and arriving at this point. It is clear that for a large
black hole these geodesics will not change very much for time
intervals for which the Hawking radiation has set in but during which only
a neglible fraction of the black hole mass has been radiated away.  Therefore,
it would seem that an iterative self-consistent solution should be
possible.

\section{Entropy?}

While the two issues of entropy and the issue of decoupling of modes at $r=0$ are
very interesting, there is no time to discuss them in this talk.  A discussion
of these issues will appear in a forthcoming paper.


\begin{thebibliography}{10}

\bibitem{bhlett}
Kirill Melnikov and Marvin Weinstein, {\bf hep-th/0109201} (2001),

\bibitem{hawking}
S. W. Hawking, Commun. Math. Phys. {\bf 43}, 199 (1975); J.B. Hartle 
and S.W. Hawking, Phys. Rev. {\bf D13}, 2188 (1976).

\bibitem{unruh} W.G. Unruh, Phys. Rev. {\bf D14}, 870 (1976).

\bibitem{jacobson} 
T.~Jacobson and D.~Mattingly,
Phys. Rev.  {\bf D61}, 024017 (2000).

\bibitem{parikh} M. Parikh and F. Wilczek, Phys. Rev. Lett. {\bf 85}, 
5042 (2000).

\bibitem{fen}  If the initial state is not the vacuum state but 
any finite energy state of the instantaneous Hamiltonian, the 
finite energy will be radiated away in finite amount of time.
The structure of radiation at large later times is independent 
of the initial quantum state of the massless field. 


\bibitem{birel} N.D. Birrell  and P.C.W. Davies, {\it Quantum Fields
in the curved space}, Cambridge University Press, 1984.

\bibitem{dewitt79} B.S. DeWitt, in {\it General Relativity}, eds. S.W. Hawking 
and W. Israel, Cambridge University Press, 1979.

\end{thebibliography}
\end{document}